\documentclass[12pt,nohyper,notoc]{JHEP}
\usepackage{amsmath,amssymb,cite,graphicx} % use drftcite in draftmode otherwise
\usepackage{epsf}

\newcommand{\PSbox}[3]{\mbox{\rule{0in}{#3}\includegraphics{#1}\hspace{#2}}}

\title{Glueballs and AdS/CFT}

\author{John Terning\\
T-8 MS B285, Los Alamos National Lab., Los Alamos NM, 87545\\
Email: {\tt terning@lanl.gov}}

\abstract{I review the calculation of 
the  glueball spectrum in non-supersymmetric 
Yang-Mills theory (in 3 and 4 dimensions) 
using the conjectured duality between supergravity and
large $N$ gauge theories. The glueball masses are obtained by solving the
supergravity wave equations in a black hole geometry.
The masses obtained this way are in
unexpectedly good agreement with the available lattice data, and are much 
better than strong-coupling expansion results. I also show how
to use a modified version of the duality to
calculate the glueball mass spectrum with some of the Kaluza-Klein 
states of the supergravity theory decoupled from the spectrum. 
}

\preprint{{\tt hep-ph/0204012}}

\begin{document}

\section{Introduction}              
In recent years there have been remarkable advances in our understanding
of certain non-perturbative gauge theories. In particular
Maldacena's conjecture\cite{Malda} relates ${\cal N}=4$ supersymmetric
$SU(N)$ gauge theories in the large $N$ limit to Type IIB string theory
on a certain background spacetime.  This arises from the fact that $N$
D3-branes stacked on top of each other have a low-energy description as
a 4 dimensional (4D) $SU(N)$ gauge theory. D3-branes are 
membranes with 3 spatial dimensions
on which strings can end with Dirichlet boundary conditions. There are
$N^2$ ways for strings to connect $N$ D3-branes, and as the D3-branes are
brought together, the string lengths, and hence masses of the lightest modes,
go to zero. This gives a $U(N)$ gauge theory, but the $U(1)$ factor is 
infrared free, so at low-energies we are left with an $SU(N)$ gauge theory.
The D3-branes are also a  gravitational
source that warps the 10D space they live in.  Taking a low-energy limit
in this spacetime 
corresponds to approaching the horizon that forms around 
the D3-branes. In this limit the metric reduces to that of 
AdS$_5\times {\bf S}^5$, where AdS$_5$ is a 5D
anti-de Sitter (negatively curved, with a negative cosmological constant)
space and ${\bf S}^5$ is a 5D sphere. 
The metric of this space is given by
  \begin{equation}
 {ds^2 \over l_s^2 \sqrt{4 \pi g_s N}} = \rho^{-2} d\rho^2 +
\rho^2
\sum_{i=1}^4 dx_i^2 + d \Omega_5^2
\label{ads5}
\end{equation}
where $l_s$ is the string length related to the superstring
tension, $g_s$ is the string coupling constant
and $d\Omega_5$ is the line element on ${\bf S}^5$. The boundary of
AdS$_5$ at $\rho=\infty$
is a flat 4D space, ${\bf R}^4$, where we take the
corresponding gauge theory to live. The
$x_{1,2,3,4}$ directions in AdS$_5$ are also the coordinates
on the ${\bf R}^4$. The gauge coupling constant $g_4$ of
the 4D theory is related to the string
coupling constant $g_s$ by $g_4^2 = g_s$.
In the 't Hooft limit
($N \rightarrow \infty$ with $g_4^2N$ fixed),
the string coupling constant vanishes: $g_s \rightarrow 0$.
Therefore one can study the 4D gauge theory using
first quantized string theory in the AdS space
(\ref{ads5}). Moreover if $g_s N \gg 1$, the curvature
of the AdS space is small and the string theory is approximated
by classical supergravity. At first glance it is surprising (to say the least) that these two
theories (a 4D gauge theory and 10D supergravity)
could be related in any way.  However as a simple first check one
can see that the $SO(2,4)$ isometry of AdS$_5$ corresponds precisely to
the conformal symmetry of the supersymmetric gauge theory, while
the $SO(6)$ isometry of  ${\bf S}^5$ corresponds to the  $SO(6)$
global symmetry of the  supersymmetric gauge theory (which has 6 scalars
and 4 fermions in the adjoint representation of the gauge group).

Remarkably
Witten has extended this proposed correspondence
to non-supersymmetric theories\cite{witten}. In his setup 
supersymmetry is broken by heating up the gauge theory, which
corresponds to putting the 4D theory on a circle and
assigning anti-periodic boundary conditions to the fermions. In this case
the fermions get a supersymmetry breaking mass term of the order
$T=1/(2\pi R)$, where $R$ is the radius of the compact coordinate and $T$ is
the corresponding temperature, while the scalars (not protected by
supersymmetry anymore) get masses from loop corrections. Thus in
the $T\to \infty$ limit this should reproduce a pure 3D
$SU(N)$ theory,  QCD$_3$,
in the large $N$ limit. On the 
string theory side this corresponds to replacing the AdS metric
by an AdS-Schwarzschild metric describing a black hole with 
Hawking temperature
$T$ in an asymptotically AdS 
space. This metric is given by

\begin{equation}
  {ds^2 \over l_s^2 \sqrt{4 \pi g_s N}}
=
\left(\rho^2 - {b^4 \over \rho^2}
\right)^{-1} d\rho^2 +
\left(\rho^2 - {b^4 \over \rho^2}
\right) d \tau^2 + \rho^2
\sum_{i=1}^3 dx_i^2 + d\Omega_5^2,
\label{metric}
\end{equation}
where $\tau$ parameterizes the compactified circle and
the $x_{1,2,3}$ directions correspond to the ${\bf R}^3$
where QCD$_3$ lives. The horizon of this geometry is
located at $\rho=b$ with
\begin{equation}
    b = {1 \over 2R}=\pi T.
\label{horizonlocation}
\end{equation}

It is worthwhile to consider how confinement arises in such a theory.
Imagine adding an infinitely heavy quark anti-quark pair in the theory
with $T=0$. In
the string theory description they sit at the boundary $\rho=\infty$
and are connected by a string. This string will minimize it's energy
by following a geodesic between the two end points.In this
AdS space  however the geodesic is not a line with $\rho=\infty$; 
the string bends down into the interior, because
the space is warped . The further
the quarks are apart the closer the string gets to $\rho=0$. If
the quarks are a distance $r$ apart as measured on the 3D boundary,
then we can easily calculate the potential energy of the configuration, $V(r)$.
When the quark anti-quark pair are infinitely far apart the
string stretches radially inward to $\rho=0$ and we can think of this
piece of the 
energy as a renormalization of the quark constituent mass.
It is a simple geometric exercise to show that
\begin{equation}
V(r)-V(\infty)\propto \frac{1}{r}~.
\label{coulpotential}
\end{equation}
This Coulombic potential is exactly what we expect in the
$T=0$ theory since it is supersymmetric and not confining.
Turning to the non-supersymmetric theory with $T\ne0$ we see
that the string cannot bend arbitrarily far into the interior,
an external observer will never see it fall through the horizon of
the black hole, so for large $r$ the string will
stretch down towards the horizon at $\rho=b$, 
and since the length of the string along
the horizon is just proportional to $r$ we see that the long distance
potential is confining\cite{witten}:
\begin{equation}
V(r)-V(\infty)\propto r~.
\label{potential}
\end{equation}
Thus the correspondence gives us a simple way to think about confinement:
the existence of the black hole horizon in the supergravity description
corresponds to confinement in the gauge theory description.

The supergravity approximation is valid for this theory when
the curvature of the space is small, thus when $g_s N\to \infty$. 
However, in order to obtain the pure 3D gauge theory  
one must take the limit $T
\rightarrow \infty$. In order to keep the intrinsic
scale $g_3^2 N =g_4^2 N/R$ of the resulting theory fixed, 
one also needs to take $g_4^2 N=g_s N \to 0$. (Here $g_3$ is 
the dimensionful gauge coupling of  QCD$_3$.)
This is exactly the opposite limit in which the supergravity approximation
is applicable! Thus, as expected for any strong-weak duality, 
the weakly coupled classical supergravity theory and the QCD$_3$ theory
are valid in different limits of the 't Hooft coupling $g_4^2 N$. 

From the point of view of QCD$_3$, the radius $R$ of the
compactifying circle provides  an ultraviolet cutoff
scale. 
Therefore, the straightforward application of 
the Maldacena conjecture
can only be used to study large $N$ QCD with a fixed ultraviolet
cutoff $R^{-1}$ in the strong ultraviolet coupling regime.
One must hope that the results
one obtains this way are not very sensitive to removing the cutoff, that is
on going from one limit to the other.
Since the theory is non-supersymmetric, there is a priori no reason 
to believe that these two limits have anything to do with each other, since
for example there might very well be a phase transition when the
't Hooft coupling is decreased from the very large values where the 
supergravity description is valid to the small values where the theory 
should describe QCD$_3$. 
Nevertheless, Witten showed that the supergravity
theory correctly reproduces several of the qualitative features of a confining
3D pure gauge theory correctly\cite{witten}. As we have 
already seen
it produces area-law confinement, in addition he showed that there is a
mass gap in the spectrum. Here I will address the question of whether
any of the quantitative features of the gauge theories are reproduced as 
well. In particular, I will review the 
calculation of the glueball mass spectrum\cite{COOT}, and show 
that it is in reasonable agreement with recent lattice
simulations\cite{Teper,Morningstar}.

\section{The 3D Glueball Spectrum}

The first step in the calculation is to identify the
operators in QCD$_3$ that have quantum numbers
corresponding to the glueball of interest. For example, 
the lowest dimension 
operator with quantum numbers $J^{PC}=0^{++}$ is
${\rm Tr} F^2 \equiv {\rm Tr} F_{\mu \nu}F^{\mu \nu}$. 
According to the refinement of the
Maldacena conjecture given in\cite{ref}, for each  supergravity
mode there is a corresponding operator in the gauge theory; this operator
couples to the supergravity 
mode on the boundary of the AdS space. To calculate the  $0^{++}$ glueball 
mass spectrum one could  evaluate the
correlators $\langle{\rm Tr} F^2(x)  
{\rm Tr} F^2(y) \rangle =
\sum_i c_i e^{-m_i |x-y|}$, and extract the $m_i$'s 
which are the glueball masses.
However the masses can be obtained by simply
solving the supergravity wave equations for
the modes that couple to these operators on the boundary. In the case
of the $0^{++}$ glueballs, one needs to find the solutions of the
dilaton equations of motion, since it is the dilaton that couples to 
${\rm Tr} F^2$.
In the supergravity theory on AdS$_5\times {\bf S}^5$, the
Kaluza-Klein (KK) modes on the ${\bf S}^5$ can be classified according to the 
spherical harmonics of the ${\bf S}^5$, which form representations of
the isometry group $SO(6)$. The states carrying non-trivial $SO(6)$ 
quantum numbers are heavier and
do not couple to pure gluonic operators on the boundary (since
the gluons are singlets under $SO(6)$),
thus the glueballs are identified with
the $SO(6)$ singlet states. Therefore we consider
dilaton solutions propagating on the 
boundary of AdS with 3D momentum $k$ which
have the form $\Phi =f(\rho ) e^{ikx}$. 
One looks for normalizable regular solutions to the dilaton
equation of motion which will give a discrete spectrum\cite{witten} with the
glueball masses determined by the eigenvalues $M_i^2=-k_i^2$.

In the supergravity description one has to solve the classical
equation of motion of the massless dilaton,
\begin{equation}
\label{dilaton}
\partial_{\mu} \left[ \sqrt{g} \partial_{\nu}\Phi
g^{\mu \nu} \right] =0 \ ,
\end{equation}
on the AdS$_5$ black hole background (\ref{metric}).
Plugging the ansatz $\Phi =f(\rho )e^{ikx}$ into this 
equation and using the metric
(\ref{metric}) one obtains the following differential equation for
$f$:
\begin{equation}
\rho^{-1} {{d}\over{d \rho}} \left( \left(\rho^4 -
b^4 \right) \rho {{d f}\over{d \rho}} \right) - k^2 f = 0
\label{dilatondiff}
\end{equation}
In the following I set $b=1$, so the masses are computed
in units of $b$.
The task
is to solve this equation as an eigenvalue problem for $k^2$. 
For large $\rho$, the black hole
metric (\ref{metric}) asymptotically approaches
the AdS metric, and the
behavior of the solution for a $p$-form
for large $\rho$ takes the form  $\rho^{\lambda}$, where $\lambda$ is
determined from the mass $M$ of the supergravity field:
\begin{equation}
  M^2 = \lambda( \lambda + 4 - 2p)~.
\label{power}
\end{equation}
Indeed for the dilaton ($M=0$, $p=0$) 
both (\ref{dilatondiff}) and (\ref{power}) give the
asymptotic forms $f \sim 1, \rho^{-4}$, and
only the later is a normalizable solution\cite{witten}.
Since the black hole geometry is regular at the horizon
$\rho=1$, $k^2$ has to be adjusted so that
$f$ is also regular\cite{witten} at $\rho=1$. 
This can be done numerically in a 
simple fashion using 
a ``shooting" technique as follows.
For a given value of $k^2$ the equation is numerically integrated from
some sufficiently
large value of $\rho$ ($\rho \gg k^2$) by matching
$f(\rho)$ with the asymptotic
solution.
The results obtained this way, 
together with the results of the lattice simulations\cite{Teper}
are displayed in Table~\ref{summary}. 
Since the lattice results are
in units of string tension, I normalized the supergravity
results  so that the lightest $0^{++}$ state agrees with
the lattice result.
 One should also expect a systematic error in addition to the  statistical
error
denoted in Table \ref{summary} for the lattice computations.
Similar numerical results have been obtained by other authors\cite{jev}, 
while a WKB approximation for the eigenvalues of (\ref{dilatondiff}) has 
been 
obtained by Minahan\cite{Minahan}.
% TABLE 1
\begin{table}[htbp]
\centering
\caption{$0^{++}$ glueball masses in QCD$_3$
coupled to ${\rm Tr}~F_{\mu \nu}
F^{\mu\nu}$. The lattice results are in units of the square root
of the string tension.\label{summary}}
\begin{tabular}{|l|ccc|}
\hline
state & lattice, $N=3$ & lattice, $N\rightarrow \infty$ &
supergravity  \\
 \hline
 $0^{++}$ & $4.329 \pm 0.041$ & $4.065 \pm 0.055$ & 4.07 ({\rm input}) \\
 $0^{++*}$ & $6.52 \pm 0.09$ & $6.18 \pm 0.13$ & 7.02 \\
 $0^{++**}$ & $8.23 \pm 0.17$ & $7.99 \pm 0.22$ & 9.92 \\
 $0^{++***}$ &  - & - & 12.80 \\
\hline
\end{tabular}
\end{table}

The $0^{--}$ glueballs can be dealt with similarly by
considering  the two-form field of the supergravity theory\cite{COOT}.
Since the supergravity method and lattice gauge theory
compute the glueball masses in different units,
one cannot compare the absolute values of the
glueball masses obtained using these methods.
However it makes sense to compare the ratios of glueball
masses.
It turns out that the supergravity
results
are in good agreement with the lattice gauge theory
computation\cite{Teper}, for example:
\begin{eqnarray}
&\left(\frac{M_{0^{--}}}{M_{0^{++}}}\right)_{{\rm supergravity}}&= 1.50 
\nonumber \\
&\left(\frac{M_{0^{--}}}{M_{0^{++}}}\right)_{{\rm lattice~~~~~}}& =
 1.45\pm 0.08 ~.
\end{eqnarray}

The results obtained from the supergravity
calculation are in reasonable agreement with the lattice results, even though
these two calculations are in the 
opposite limits for  the 't Hooft coupling. Therefore, it is 
important to see, how the ratios are modified once finite $N$
corrections (string theory corrections in the supergravity language)
are taken into account. The leading  string theory corrections
can be calculated by using the results of Gubser, Klebanov, and 
Tseytlin\cite{GKT}, who calculated the
first stringy corrections to the AdS$_5$ black-hole metric (\ref{metric}). 
The details of the calculation can be found in\cite{COOT}, where
it was shown  that the string theory corrections are somewhat uniform for the
different excited states of the $0^{++}$ glueball, and therefore 
one could hope that these corrections to the ratios of the 
glueball masses are small. However, it can be seen that this is probably
too optimistic an assumption, by considering the KK partners of
the glueball states. The KK modes do not correspond to any state in
QCD, but rather they should decouple in the $R\to 0, g_4^2N\to 0$
limit. However, in the supergravity limit of finite
$R$, $g_4^2N \to \infty$ these states have masses comparable
to the light glueballs\cite{ORT}. This is simply a consequence of
the fact, that the masses of the fermions and scalars 
is of the order of the temperature $T$,
thus their bound states are expected to also have masses of the order of
the temperature. However,  the temperature is the only scale in 
the theory,  so 
this will also be the cutoff scale of the QCD theory, and thus
the mass scale for the glueballs. This
situation is clearly unsatisfactory. One may try to 
improve on it by introducing a different supergravity background, where 
some of these KK modes are automatically decoupled. I will consider this
possibility in the next section where I discuss the construction based
on rotating branes\cite{Russo,CORT,CRST}.

\section{The 4D Glueball Spectrum}

Results similar to the the ones presented in the previous section can be
obtained for the glueball mass spectrum in QCD$_4$ by starting 
from a slightly different construction where the M-theory 5-brane 
is wrapped on two circles\cite{witten}. The details of these results can be
found in\cite{COOT}. Here I will review only the generalized
construction based on the rotating M5 brane with one angular momentum,
first constructed in\cite{Russo}, and explored in\cite{CORT}.
The metric for this background is given by
\begin{eqnarray}
\label{pocho}
ds^2_{\rm IIA}={2\pi \lambda A \over  3u_0} u \Delta ^{1/2}\bigg[&& 4u^2
\big( -dx_0^2+dx_1^2+dx_2^2+dx^2_3\big)
+ { 4A^2\over 9u_0^2} u^2 \ (1-{u_0^6\over u^6 \Delta }) d\theta_2^2 
\nonumber \\
&&+ {4\ du^2  \over u^2 (1-{a^4\over u^4}-{u_0^6\over u^6 }) }
+ d\theta^2+{\tilde{\Delta}\over \Delta} \sin^2\theta d\varphi^2 \\
&&+{1\over \Delta } \cos^2\theta d\Omega_2^2
 -{4a^2 A u_0^2\over 3u^4\Delta } \sin^2\theta d\theta_2 d\varphi \bigg], \nonumber 
\end{eqnarray}
where $x_{0,1,2,3}$ are the coordinates along the  brane where the
gauge theory lives, $u$ is the ``radial" coordinate of the AdS space, while the
remaining four coordinates parameterize the angular variables of $S^4$, 
$a$ is the angular momentum parameter, and I have introduced
\begin{equation}
\Delta=1-{a^4\cos^2\theta \over u^4}\ ,\ \ \ \ \tilde{\Delta}=1-{a^4\over 
u^4} \ ,
\ \ \ \ A\equiv {u_0^4\over u_H^4-\frac{1}{3} a^4}\ , \ \ \ \ 
u_H^6-a^4 u_H^2-u_0^6=0\ .
\end{equation}
$u_H$ is the location of the horizon, and the dilaton background 
and the temperature of the field theory are  given by
\begin{eqnarray}
e^{2\Phi }={8\pi\over 27} {A^3\lambda^3 u^3\Delta^{1/2}\over u_0^3} 
{1\over N^2}\ , \ \ \ \  R=(2\pi T_H)^{-1}={A\over 3u_0}\ .
\label{dilz}
\end{eqnarray}
Note, that in the limit when $a/u_0\gg 1$, the radius of compactification $R$
shrinks to zero, thus the KK modes on this compact direction are 
expected to decouple in this theory when the angular momentum $a$ is increased.
In order to find the mass spectrum of the $0^{++}$ glueballs, one needs
to again solve the dilaton equations of motion as a function of $a$. 
This can be done by plugging the background (\ref{pocho}) into the 
dilaton equation of motion 
\begin{equation}
\partial_{\mu} \left[ \sqrt{g} e^{-2 \Phi} 
g^{\mu \nu} \partial_{\nu} \Phi \right]=0.
\end{equation}
This can be solved as explained in the previous section,
where the eigenvalues are now a function of the angular momentum 
parameter $a$.
The results of this are summarized in Table \ref{tab:4ddila}. 
Note, that while some
of the KK modes decouple in the $a\to \infty$ limit, the 
$0^{++}$ glueball mass ratios change only very slightly, showing that the
supergravity predictions are robust for these ratios against the change
of the angular momentum parameter.

\begin{table}[htbp]
\centering
\caption{Masses of the first few $0^{++}$ glueballs in
QCD$_4$, in GeV. The change
from $a=0$ to $a=\infty$  is
tiny.
\label{tab:4ddila}}
\begin{tabular}{|l|ccc|}
\hline
state & lattice, $N=3$ &
supergravity  $a=0$ & supergravity $a\to \infty$ \\
 \hline
 $0^{++}$ & $1.61 \pm 0.15$   & 1.61 {\rm (input)} & 1.61 {\rm (input)} \\
 $0^{++*}$ &  $2.48 \pm 0.23 $  & 2.55 &  2.56 \\
 $0^{++**}$ &   - & 3.46  & 3.48 \\
\hline
\end{tabular}
\end{table}

One can similarly calculate the mass ratios for the $0^{-+}$ glueballs,
by considering the equations of motion of the supergravity mode
that couples to
the operator ${\rm Tr}F\tilde{F}$. To find the $0^{-+}$
glueball spectrum one has 
to solve the supergravity equation of motion:
\begin{equation}
\partial_{\nu} \left[ \sqrt{g} g^{\mu\rho}g^{\nu\sigma}
(\partial_{\rho}A_{\sigma}-\partial_{\sigma}A_{\rho})\right]=0
\end{equation}
in the background (\ref{pocho}). 
The results are summarized in Table \ref{tab:0-+}. Note, that the
change in the $0^{-+}$ glueball mass is sizeable when going from $a=0$ to
$a\to \infty$, and is in the right direction as suggested by lattice 
results\cite{Teper,Morningstar}.

\begin{table}[htbp]
\centering
\caption{Masses of the first few $0^{-+}$ glueballs in
QCD$_4$, in GeV. Note that the change
from $a=0$ to $a=\infty$ in the supergravity predictions is 
of the order $\sim 25 \%$.\label{tab:0-+}}
\begin{tabular}{|l|ccc|}
\hline
state & lattice, $N=3$ &
supergravity  $a=0$ & supergravity $a\to \infty$ \\
 \hline
 $0^{-+}$ & 2.59 $\pm$0.13   & 2.00 & 2.56 \\
 $0^{-+*}$ &  3.64 $\pm$0.18   & 2.98 &  3.49 \\
 $0^{-+**}$ &   - & 3.91  & 4.40 \\
\hline
\end{tabular}
\end{table}

One can also calculate the masses of the different KK modes
in the background of (\ref{pocho}). One finds, that as expected from the fact
that for $a\to \infty$ the compact circle shrinks to zero, the KK modes
on this compact circle decouple from the spectrum, leading to  a   
real 4D gauge theory in this limit. However, the KK modes 
of the sphere $S^4$ do not decouple from the spectrum 
even in the $a\to \infty$ limit. These conclusions remain unchanged even
in the case when one considers the theory with the maximal number of
angular momenta (which is two for the case of QCD$_4$) 
\cite{CRST,JorgeKostas}. 
In the limit when the angular momentum becomes large, $a/u_0\gg 1$,
the theory approaches a supersymmetric limit\cite{Russo,CRST}
since the supersymmetry breaking fermion masses get smaller
with increasing angular momentum\cite{CG}. Therefore, the limit of
increasing angular momentum on one hand does decouple some of the
KK modes which makes the theory four dimensional, but at the same time
re-introduces the light fermions into the spectrum\cite{CG}. 
 
The final results for the 4D spectrum are shown in Figure 1, compared to the
lattice results\cite{Teper,Morningstar}.  The results
are much better than we have any (known) reason to expect: they are within 4\%
of the lattice results.  This can be contrasted with the classic
predictions of the strong-coupling expansion\cite{Kogut} 
(with the simplest Wilson action) which are off by
between 7\% and 28\%.
\begin{figure}[ht]
\PSbox{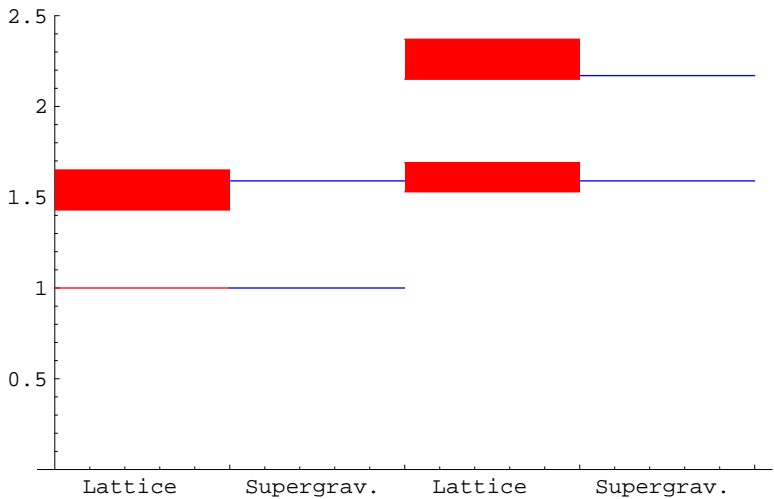 hscale=85 vscale=85 hoffset=-60
voffset=-430}{7cm}{6cm}
%%%%\figurebox{20pc}{15pc}{} % to have a box alone
%\epsfxsize=23pc 
% will enlarge or reduce the postscript figures based on the xsize
%\epsfbox{spectrum.eps} % postscript image file name
\caption{Lattice and supergravity estimates of 
glueball masses divided by the mass of the $0^{++}$.  Shown on the
left are the first two  $0^{++}$ states; on the
right the first two $0^{-+}$ states.  
\label{fig:spectrum}}
\end{figure}

\section{Conclusions}
I have described how Maldacena's conjecture can be used
to study pure Yang-Mills theories in the large $N$ limit. 
These methods reproduce several of the qualitative features of QCD,
and, in addition, one finds that the supergravity calculations are 
in a reasonable agreement with the lattice results, even though they are 
obtained in the opposing limits
of the 't Hooft coupling. It would be very important to understand whether
this unexpected agreement is purely a numerical coincidence or whether there
is any deeper reason behind it. The historical 
precedent for a deeper explanation for unexpectedly good results in QCD
phenomenology was of course
discussed in detail at this conference: some of the successes of the quark
model and the Skyrme model are explained in the large $N$ limit!

\section*{Acknowledgments}
I thank Csaba Cs\'aki,
H. Ooguri, Y. Oz, J. Russo and K. Sfetsos for
several collaborations which this talk is based on. 
I thank M. Strassler, D.T. Son, and M. Teper for useful discussions
during the conference.
This work was supported in part
the U.S. DOE under contracts DE-AC03-76SF00098 and
W-7405-ENG-36; and in part
by the NSF under grant PHY-95-14797.

\end{document}